\documentclass[twocolumn]{aastex701}

\definecolor{mylinkcolor}{RGB}{50,68,135}

\shorttitle{29~Cyg~b JWST Coronagraphy}
\shortauthors{W. O. Balmer et al.}

\hypersetup{linkcolor=mylinkcolor,citecolor=mylinkcolor,urlcolor=mylinkcolor}

\usepackage{xspace}
\newcommand{\um}{\text{\,\textmu m}\xspace}
\accepted{Jan. 13th, 2026}

\submitjournal{The Astrophysical Journal Letters}

\begin{document}

\title{Direct Images of CO$_2$ Absorption in the Atmosphere of a Super-Jupiter:\\Enhanced Metallicity Suggestive of Formation in a Disk}
% \title{Living on the Wedge II:\\A Super-Jupiter With Enhanced Metallicity Suggestive of Formation in a Disk}

\author[0000-0001-6396-8439,sname=Balmer]{William O. Balmer}
\email[show]{wbalmer1@jhu.edu}
\affiliation{Department of Physics \& Astronomy, Johns Hopkins University, 3400 N. Charles Street, Baltimore, MD 21218, USA}
\affiliation{Space Telescope Science Institute, 3700 San Martin Drive, Baltimore, MD 21218, USA}
\correspondingauthor{W. O. Balmer}

\author[0000-0003-3818-408X]{Laurent Pueyo}
\email{pueyo@stsci.edu}
\affiliation{Space Telescope Science Institute, 3700 San Martin Drive, Baltimore, MD 21218, USA}

\author[0009-0002-4970-3930]{Ashley Messier}
\email{amessie2@jh.edu}
\affiliation{Department of Physics \& Astronomy, Johns Hopkins University, 3400 N. Charles Street, Baltimore, MD 21218, USA}

\author[0009-0001-0275-7811]{Evelyn Bruinsma}
\email{ebruins1@jhu.edu}
\affiliation{Department of Physics \& Astronomy, Johns Hopkins University, 3400 N. Charles Street, Baltimore, MD 21218, USA}

\author[0000-0003-3045-5148]{Jeremy Jones}
\email{jjones176@gsu.edu}
\affiliation{Center for High Angular Resolution Astronomy and Department of Physics and Astronomy, Georgia State University, 25 Park Place, Suite 605, Atlanta, GA, 30303, USA}

\author[0009-0007-3572-0664]{Klara Matuszewska}
\email{kmatuszewska26@amherst.edu}
\affiliation{Amherst College, Department of Physics and Astronomy, Department of Physics \& Astronomy, Amherst College, 25 East Drive, Amherst, MA 01002, USA}

\author[0000-0002-3191-8151]{Marshall D. Perrin}
\affiliation{Space Telescope Science Institute, 3700 San Martin Drive, Baltimore, MD 21218, USA}
\email{mperrin@stsci.edu}

\author[0000-0001-8627-0404]{Julien H. Girard}
\affiliation{Space Telescope Science Institute, 3700 San Martin Drive, Baltimore, MD 21218, USA}
\email{jgirard@stsci.edu}

\author[0000-0002-0834-6140]{Jarron M. Leisenring}
\affiliation{Steward Observatory and Department of Astronomy, University of Arizona, 933 N Cherry Ave, Tucson, AZ 85721, USA}
\email{jarronl@arizona.edu}

\author[0000-0002-6964-8732]{Kellen Lawson}
\email{kellen.d.lawson@nasa.gov}
\affiliation{Center for Space Sciences and Technology, University of Maryland, Baltimore County, 1000 Hilltop Circle, Baltimore, MD 21250, USA}
\affiliation{Astrophysics Science Division, NASA-GSFC, 8800 Greenbelt Rd, Greenbelt, MD 20771, USA}
\affiliation{Center for Research and Exploration in Space Science and Technology, NASA-GSFC, 8800 Greenbelt Rd, Greenbelt, MD 20771, USA}

\author[0000-0001-7827-7825]{Roeland P. van der Marel}
\affiliation{Space Telescope Science Institute, 3700 San Martin Drive, Baltimore, MD 21218, USA}
\affiliation{Department of Physics \& Astronomy, Johns Hopkins University, 3400 N. Charles Street, Baltimore, MD 21218, USA}
\email{marel@stsci.edu}

\author[0000-0003-2769-0438]{Jens Kammerer}
\email{jkammere@eso.org}
\affiliation{European Southern Observatory, Karl-Schwarzschild-Stra\ss e 2, 85748 Garching, Germany}

\author[0000-0001-5365-4815]{Aarynn Carter} 
\email{aacarter@stsci.edu}
\affiliation{Space Telescope Science Institute, 3700 San Martin Drive, Baltimore, MD 21218, USA}

\author[0000-0002-2918-8479]{Mathilde Mâlin}
\affiliation{Department of Physics \& Astronomy, Johns Hopkins University, 3400 N. Charles Street, Baltimore, MD 21218, USA}
\affiliation{Space Telescope Science Institute, 3700 San Martin Drive, Baltimore, MD 21218, USA}
\email{mmalin@stsci.edu}

\author[0000-0002-4479-8291]{Kimberly Ward-Duong}
\email{kwardduong@smith.edu}
\affiliation{Department of Astronomy, Smith College, Northampton, MA, 01063, USA}

\author[0000-0002-9803-8255]{Kielan K. W. Hoch}
\email{khoch@stsci.edu}
\affiliation{Space Telescope Science Institute, 3700 San Martin Drive, Baltimore, MD 21218, USA}

\author[0000-0003-4203-9715]{Emily Rickman} 
\email{erickman@stsci.edu}
\affiliation{European Space Agency (ESA), ESA Office, Space Telescope Science Institute, 3700 San Martin Drive, Baltimore MD, 21218}

\author[0000-0002-6892-6948]{Sara Seager}
\email{seager@mit.edu}
\affiliation{Department of Earth, Atmospheric and Planetary Sciences, Massachusetts Institute of Technology, Cambridge, MA 02139, USA}
\affiliation{Department of Physics and Kavli Institute for Astrophysics and Space Research, Massachusetts Institute of Technology, Cambridge, MA 02139, USA}
\affiliation{Department of Aeronautics and Astronautics, MIT, 77 Massachusetts Avenue, Cambridge, MA 02139, USA}

% Kim, Sara

% need others to add their emails, affiliations, etc

% \author{\todo{more}}
% \email{test@gmail.com}
% \affiliation{}

\begin{abstract}

It is unclear how directly imaged substellar companions with masses near the deuterium burning limit form, because these objects are rare and their bulk properties are not diagnostic of their formation. In this paper we revisit this problem using JWST/NIRCam coronagraphic images of the 29~Cygni (=HIP~99770) system that reveal the recently-discovered super-Jovian companion 29~Cyg~b at wavelengths covering 4-5\um for the first time. This object has an uncertain mass that straddles the deuterium burning limit ($M_{\rm b}\simeq15\pm5\,M_{\rm J}$) and a low mass ratio with its early-type host star ($M_{\rm b}/M_\star\sim0.01$). Absorption from CO$_2$ and CO is apparent at 4.3 and 4.6\um in our images. The strength of the CO$_2$ feature relative to CO provides strong evidence, based on empirical comparison with literature observations at these wavelengths and atmospheric modeling, that the companion is enriched in heavier elements compared to the roughly solar abundances of the host ($Z_{\rm b}/Z_\star=3\pm2$). 
In addition, we measure the stellar inclination angle with CHARA/PAVO interferometry: the system is consistent with spin-orbit alignment at the $2\,\sigma$ level, with $\Delta i=12\pm6^\circ$. This ensemble of evidence is suggestive of formation within the protoplanetary disk and rapid accretion of metal-rich material, versus disk fragmentation or capture like higher mass ratio companions. 29~Cyg~b shows that planet formation around early-type stars can occur on scales at or exceeding the deuterium burning limit, in agreement with the recently revised planetary mass/metallicity trend that predicts $Z_{\rm pl}/Z_\star=3.3\pm0.5$ at high masses from transiting planet densities \citep{Chachan2025}.

\end{abstract}

\section{Introduction}
The formation mechanisms that produce wide separation ($10-100\,\mathrm{au}$), low mass ratio ($M_{\rm b}/M_\star\equiv q\lesssim0.01$) directly imaged companions near the deuterium burning limit ($\sim13\,M_{\rm J}$) have not been unambiguously distinguished. Observations are confounded by the small sample size and overlap in the theoretical predictions of the most massive outcomes of core accretion \citep[][]{Emsenhuber2021} and the least massive outcomes of disk instability \citep{Kratter2016} and molecular cloud fragmentation and capture \citep{Bate2009}. Mass and luminosity on their own appear insufficient to distinguish between formation pathways \citep{Molliere2012, Mordasini2013, Brandt2014}. Nevertheless, distinguishing between these pathways is important, because these rare, young, massive planets provide unique insight into the initial conditions of giant planet formation \citep{Marleau2014}. In principle, formation within a protoplanetary disk encodes varying abundances of heavy elements in the planet's atmosphere, which are different from the host star's composition \citep{Oberg2011, Molliere2022} and could be used to diagnose formation history. In reality, these measurements have been extremely challenging to make and interpret, because of the complicated, non-Gaussian and telluric noise inherited from high contrast, ground based observations, the small signal amplitude of composition dependent atmospheric features in the near infrared, and uncertain bulk properties like mass, radius, and age for these systems \citep[see the discussion in, e.g,][]{Feinstein2025}. 
\par Coupling absolute astrometric measurements from Hipparcos and Gaia with high contrast imagers has recently identified new sub-stellar companions with dynamical mass constraints \citep[e.g.,][]{Bonavita2022, Franson2023, Li2023}, whose atmospheres can be more comprehensively interpreted with strong a-priori information about their surface gravity. Nevertheless, deficiencies and degeneracies in data/model inference can produce unphysical results regardless of these powerful priors. These mass measurements alone have not been able to resolve the many degeneracies between surface gravity, cloud properties, and metallicity that preclude unambiguous abundance determination in the near-infrared \textit{J}, \textit{H}, and \textit{K}-bands from noisy data \citep[see discussion in, e.g.,][]{Balmer2023, Balmer2025a}. Despite the challenge of accurately measuring a directly imaged companion's metallicity, there is a slowly growing consensus that the atmospheric properties of companions with mass ratios $M_{\rm p}/M_\star\lesssim0.01$ exhibit an apparently large enrichment compared to their host stars \citep{Molliere2020, Zhang2023, Nasedkin2024} and that these companions may have accreted a significant amount of solids at early ages \citep{Wang2025}. 

\citet{Chachan2025} recently used transiting warm giant planets that have measured masses and radii to revise the planetary mass/metallicity trend \citep{Guillot2006, Miller2011, Thorngren2016}. They show that this relationship plateaus to an enriched value $Z_{\rm p}/Z_\star=3.3\pm0.5$ at high masses ($10-20\,M_{\rm J}$), indicating that giant planets accrete enriched material during their gas accretion phase, in contradiction with the classical core accretion framework where metal-rich material is accreted solely during core formation, which is then diluted by metal-poor gas during runaway accretion. Do giant planets accrete metal-rich solids or metal-rich gas during the gas accretion phase? What kind of higher order physical processes in the circumplanetary disk can produce the larged observed metallicities? Are these observations compatible with formation by gravitational instability, or do they require the assembly of solid cores? These new questions require precise atmospheric analysis to provide accurate elemental and isotopic abundances to test \citep{Bergin2024}.

\par JWST has dramatically expanded the study of giant planet atmospheres, thanks to its access to molecules like CO$_2$ at 4.3\um that are sensitive tracers of atmospheric metallicity \citep[e.g.][their Figure 13]{Rustamkulov2023}, but are present in the Earth's atmosphere and therefore difficult to measure from the ground. Coronagraphic observations of the massive benchmark Brown Dwarf HD~19467~B detected weak CO$_2$ absorption indicative of a stellar metallicity using the F410M, F430M, and F460M filters \citep{Greenbaum2023}, demonstrating this wavelength range's ability to break degeneracies between vertical mixing (or clouds) and metallicity. Observations of the HR~8799 system revealed strong CO$_2$ absorption on each of the four giant planets \citep{Balmer2025b}. %This measurement was facilitated by the novel use of the Near Infrared Camera (NIRCam) Long Wavelength Bar (\texttt{MASKLWB}) coronagraph. Placing the target star behind the most narrow point of this coronagraph allows for the detection of the continuum flux of close-in ($\rho=2-4\,\lambda/D$) directly imaged planets from 3-5\um. % WB: will introduce this in S2

\par The relative strength of CO$_2$ and CO absorption in the HR~8799 planets' atmospheres follow a track significantly distinct from the more massive field Brown Dwarfs at similar temperatures. We interpret this trend as uniform enrichment in metallicity compared to their host star.
%but vary from one another in their absolute colors. 
This result validated previous ground based studies that had found evidence for enriched metallicities in this system, albeit with a wide variety of degeneracies between chemistry and clouds \citep{Nasedkin2024}. As these were the first observations of CO$_2$ on directly imaged planets, it was unclear whether other directly imaged companions would exhibit strong CO$_2$ absorption, and whether they would follow the planetary mass/metallicity trend established by transiting planets \citep{Thorngren2016, Chachan2025}. Motivated by these results, our team is conducting a JWST coronagraphic imaging survey of directly imaged companions amenable to CO$_2$ measurements (JWST General Observer Program 6905, PI: Balmer), in order to determine which of these objects are consistent with atmospheric enrichment, and which are not. The four targets of our survey are all young, directly imaged companions with spectral types near the L-T transition and dynamical mass constraints $M<20\,M_{\rm J}$.

% \par Two have masses well below the deuterium burning limit: AF~Lep~b \cite[$M=3.7\pm0.5\,M_{\rm J}$,][]{Franson2023, Bonse2025, Balmer2025a}, and HD~95086~b \cite[$M=4-7\,M_{\rm J}$,][]{DeRosa2016, Wood2023}. Two have small mass ratios ($M_{\rm p}/M_\star\lesssim0.01$), but have mass estimates which straddle the deuterium burning limit and have been variously classified as brown dwarfs, planetary mass objects, or planets: HR~2562~B \citep[$M\leq18\,M_{\rm J}$,][]{Konopacky2016, StellaZhang2023} and 29~Cyg~b 
% \footnote{This target has been refereed to in the direct imaging literature as HIP~99770~b since its discovery \citep{Currie2023_sci}. We adopt the Flamsteed designation ``29~Cyg" and ``29~Cyg~b" here because of the significant literature on the stellar properties of the host listed under this name \citep[e.g][]{Slettebak1952, Sargent1965, Baschek1969, Gies1977}, following the study of $\lambda$~Boo and $\delta$~Scuti stars since the 1950s.} 
% \citep[$M=16\pm5\,M_{\rm J}$,][]{Currie2023_sci, Zhang2024, Winterhalder2025}. A key goal of the survey is to determine how closely the planetary mass objects hew to the mass/metallicity trend, and whether the low mass ratio objects extend this trend beyond the deuterium burning limit, or exhibit stellar metallicities.

\par 29~Cygni (b$^3$~Cyg, HD~192640, HIP~99770) is a nearby \citep[$\pi=24.55\pm0.09\,\mathrm{mas}$, $d=40.7\pm0.1\,\mathrm{pc}$,][]{GaiaCollaboration2022} A-type star with a storied observational past, tracing the development of theories regarding the interior structure, abundances, and formation of chemically peculiar early-type stars. The star exhibits $\delta$~Scuti-like pulsations \citep[e.g.,][]{Gies1977, Kusakin1996, Mkrtichian2007} and a Vega-like infrared excess \citep{Sadakane1986, Fajardo-Acosta1999}. It has $\lambda$~Bootis-like abundance patterns; near solar abundances of light elements (C, N, O, etc) but significantly depleted iron-peak elements \citep[][]{Slettebak1952, Burbidge1956, Baschek1969, Baschek1984, Venn1990, Heiter1998, Paunzen1999a, Paunzen1999b, Kamp2001}. These abundances are not thought to be reflective of the star's internal composition. In young A-stars iron-peak element depletion is theorized to be due to accretion of metal poor gas from the circumstellar disk \citep{Murphy2017}. In this view, the heavier (iron peak) elements in the circumstellar disk are locked in solid grains, which are prevented from accreting by radiation pressure, while the iron poor (but otherwise stellar abundance) gas accretes onto the star and is mixed on timescales of a few Myr \citep{Venn1990, Turcotte1993}.

\par The companion 29~Cyg~b (HIP~99770~b) was discovered orbiting at a separation of $\sim0\farcs5$ with Subaru SCExAO/CHARIS and Keck/NIRC2 \citep{Currie2023_sci}. The companion has subsequently been observed in the near infrared with Subaru SCExAO/CHARIS \citep{Bovie2025}, Keck/KPIC \citep{Zhang2024}, and VLTI/GRAVITY \citep{Winterhalder2025}. It orbits at $16\pm1\,\mathrm{au}$ with a low-to-moderate eccentricity $e<0.5$ and has a cloudy atmosphere of late L spectral type, similar to HR~8799~e, based on the latest in depth study \citep{Bovie2025}. It has a mass $M=16\pm5\,M_{\rm J}$ based on the star's proper motion anomaly between Hipparcos and Gaia \citep{Brandt2021}\footnote{As \citet{Bovie2025} describe, the low signal-to-noise ratio of the astrometric acceleration means that this mass measurement is prior dependent. In this work we adopt a uniform prior on the secondary mass, which given the currently available data yields central values of $15\pm5\,M_{\rm J}$. Adopting a commonly assumed log-uniform prior on the secondary mass leads to an inference of $M=13\pm5\,M_{\rm J}$. Either way, the companion's dynamical mass overlaps the deuterium burning limit at the $1\,\sigma$ level.}.

\section{Observations} \label{sec:obs}

We observed 29~Cyg and the PSF reference star 36~Cyg (HD~193369, HIP~100108) with JWST/NIRCam from UTC 2025 September 1, 08:02:33 to 12:48:30 as part of General Observer Program 6905. We used the \texttt{MASKLWB} coronagraph and the ``narrow" fiducial point override, following the strategy described by \citet{Perrin2018} and tested on-sky in \citet{Balmer2025b}. Assuming a perfect starlight subtraction, this offset position provides improved throughput on the companion compared to standard NIRCam observing modes (a $50\%$ transmission ``inner-working angle" of $2-4\lambda/D\sim0\farcs3$, compared to $4-6\lambda/D\sim0\farcs6$), which enables higher efficiency, multi-filter observing sequences for close-in companions. In practice, we have found that the wavefront evolution of the observatory is stable enough to warrant the adoption of this observing strategy. The wavefront drifts on the timescales between target and reference observations, the stochastic target acquisition errors, and the impact/mitigation of pupil shear, are generally far better in flight \citep{Girard2022, Rigby2023, Telfer2024} than originally assumed during the manufacture of the coronagraphic masks \citep{Krist2009, Krist2010, Mao2011}. We observed the system in four filters, the F210M, F410M, F430M, and F460M filters centered at 2.1\um, 4.1\um, 4.3\um, and 4.6\um, respectively. The companion 29~Cyg~b was detected with a signal-to-noise ratio $>5$ in every filter. The observation planning, data reduction, and source extraction is described in Appendix \ref{app:data}. The starlight subtracted images in each filter are shown in Figure \ref{fig:detections}, and the associated astrometry and photometry is recorded in Table \ref{tab:bka}.

\section{Analysis}

\subsection{Photometry and atmospheric modeling}

\begin{figure*}
    \centering
    \gridline{
    \fig{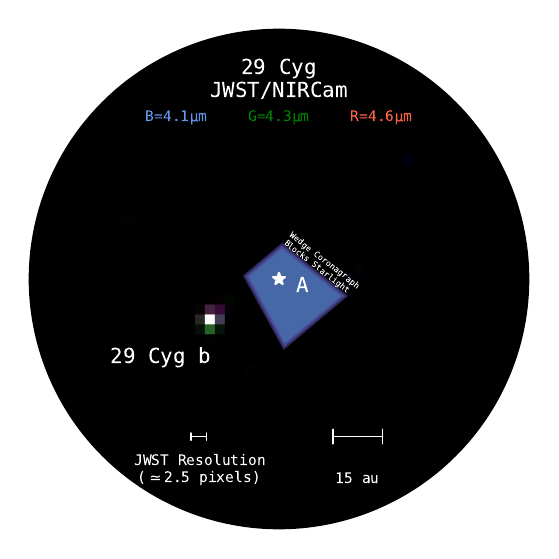}{0.5\linewidth}{(a)}
    \fig{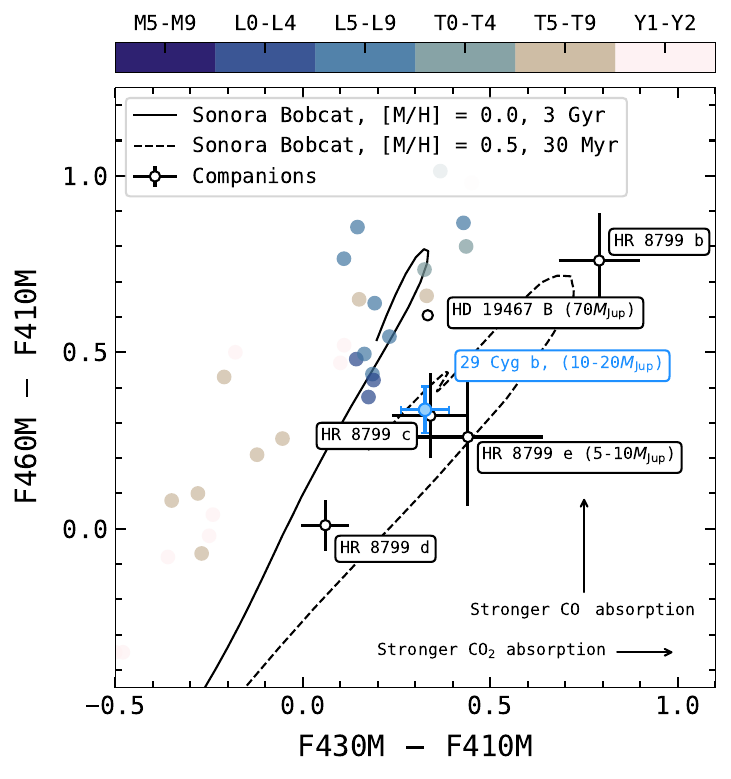}{0.5\linewidth}{(b)}
    }
    \caption{The 4-5\um color of 29~Cyg~b in context. \textbf{(a)} A deconvolved and smoothed three color composite. Blue is assigned to the F410M filter, green to the F430M filter, and red to the F460M in increasing wavelength; the brightness in each filter is in units of MJy/sr, normalized to the brightness of the companion in F410M. The location of 29~Cyg~A, blocked by the \texttt{LWBAR}/narrow coronagraph, is marked by a cartoon star. \textbf{(b)} Color-color diagram of brown dwarfs and giant planets between 4-5\um. 29~Cyg~b exhibits striking similarity with HR~8799~c and e, which are offset towards deeper CO$_2$ absorption from the brown dwarf population. Typical uncertainty for the BDs is $0.03\,\mathrm{mag}$, smaller than the scatter points. %\lap{What is the typical error bar for the BDs?}
    }
    \label{fig:color-color}
\end{figure*}

The 4-5\um photometry of 29~Cyg~b displays clear signatures of CO$_2$ and CO absorption at 4.3\um and 4.6\um, respectively. To illustrate this, we plot a false color composite of the deconvolved images in Figure \ref{fig:color-color}. The companion appears blue-white, because the absorption in the F430M and F460M filters is of a similar order of magnitude. To better contextualize this color, we show a F410M, F430M, and F460M color-color diagram that includes the four HR~8799 planets \citep{Balmer2025b} and the massive brown dwarf companion HD~19467~B \citep{Greenbaum2023} observed with the \texttt{LWBAR}, as well as the L-T field brown dwarfs observed by the AKARI space telescope \citep{Sorahana2012, Sorahana2013} and the T-Y brown dwarfs observed with JWST \citep{Beiler2024b}. The companion displays a color similar to the metal rich HR~8799~e and c planets, offset from the solar-composition brown dwarfs.

\par We compared the spectral energy distribution of the companion to radiative-convective equilibrium (RCE) models from the \texttt{Sonora Diamondback} series \citep{Morley2024} using the \texttt{species} package \citep{Stolker2020}, and to parametric models generated using the \texttt{petitRADTRANS} \citep{Molliere2019} atmospheric retrieval module \citep{Nasedkin2024_zenodo}. We acquired the \textit{JHK} (1-2.5\um) and \textit{L}' spectrophotometry from \citet{Currie2023_sci} Data S1 and Table S4,
% the \textit{H} (1.5-1.8\um) low resolution spectra from \citet{Bovie2025}, 
and the \textit{K}-band (2-2.5\um) medium resolution spectra from \citet{Winterhalder2025}, and combined these observations with our own photometry in the F210M, F410M, F430M, and F460M filters. The RCE model interpolation and comparison followed previous work with \texttt{species} \citep[e.g.][]{Stolker2020, Balmer2024, Stolker2025}. The parametric atmospheric retrieval followed previous work on emission spectroscopy with \texttt{petitRADTRANS} closely \citep{Molliere2020, Balmer2023, Zhang2023, Nasedkin2024, Balmer2025a}. The \texttt{petitRADTRANS} model set-up is described in Appendix \ref{app:atmo}. For both the model comparisons, we set normally distributed priors on the planet's mass ($M_{\rm p}=15\pm5\,M_{\rm J}$, this work, see below, but equivalent to previous work), evolutionary radius \citep[$R=1.30\pm0.15\,R_{\rm J}$, based on the mass and system age,][]{Winterhalder2025}, and parallax $\pi= 24.546\pm0.091\,\mathrm{mas}$ \citep{GaiaCollaboration2022}.

\begin{figure*}
    \centering
    \includegraphics[width=\linewidth]{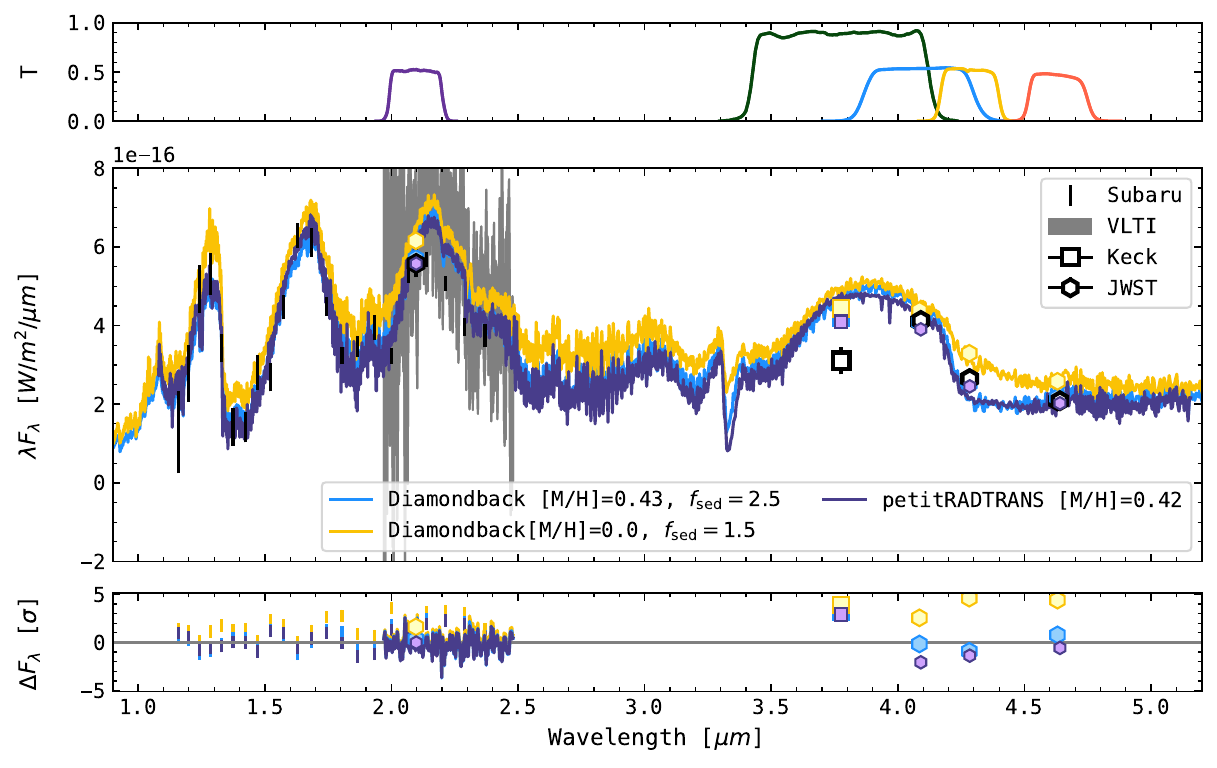}
    \caption{Comparing the spectrum of 29~Cyg~b to atmospheric models. \textit{Top:} The transmission functions of the photometry considered in the fit. \textit{Middle:} RCE models from the \texttt{Sonora Diamondback} are shown in blue and yellow, and a parametric model generated with \texttt{petitRADTRANS} is shown in purple. Observations from Subaru, the VLTI, Keck, and JWST are plotted in black and gray, and the models integrated across the filter transmission functions are shown as smaller, colored scatter points. The maximum a-posteriori spectrum (a linear interpolation across the published grid of spectra, with $\mathrm{[M/H]}=0.5$ and $f_{\rm sed}=2.5$) in blue, and a similar model with a solar metallicity ($\mathrm{[M/H]}=0.0$) and stronger cloud sedimentation ($f_{\rm sed}=1.5$) as a yellow curve in yellow. The median model has a reduced $\chi^2=1.0$. The models with enhanced metallicity in the \texttt{Diamondback} grid provide a better fit to the absorption feature from CO$_2$ at 4.3\um, while the metallicity dependent \textit{JHK}-band spectral slope could be reproduced by changing the model's cloud parameters (in this case the sedimentation efficiency, $f_{\rm sed}$). The more flexible pressure-temperature structure and cloud paramterization in the best fitting \texttt{pRT} model is able to reproduce both the \textit{K}-band observations, the \textit{JHK}-band spectral slope, and the 4-5\um photometry at once, while preserving the surface gravity and radius predicted by the dynamical mass and evolutionary models. The median \texttt{pRT} model has a reduced $\chi^2=1.0$. \textit{Bottom:} Residuals (in units of standard deviation) between the models and the observations as a function of wavelength.
    }
    \label{fig:spectrum}
\end{figure*}

The comparison with the RCE \texttt{Sonora Diamondback} models found $T_{\rm eff}=1250\pm10\,\mathrm{K}$, $\log{g}=3.7\pm0.1$, $R_{\rm p}=1.2\pm0.05\,R_{\rm J}$, (resulting in an inferred mass of $M=2.8\pm0.8\,M_{\rm J}$), $\mathrm{[M/H]}=0.4^{+0.1}_{-0.1}$, and $f_{\rm sed}=2.5\pm0.1$. This model is visualized in Figure \ref{fig:spectrum}. The sampler converged to the edge of the grid at $\mathrm{[M/H]}=0.4-0.5$, and produced unphysical surface gravities despite the dynamical mass prior, which motivated our analysis with the parametric retrieval.

The atmospheric retrieval found $T_{\rm eff}=1300\pm30\,\mathrm{K}$, $\log{g}=4.5\pm0.2$, $R_{\rm p}=1.20\pm0.05\,R_{\rm J}$, $\mathrm{[M/H]}=0.45^{+0.25}_{-0.35}$, $\mathrm{C/O}=0.45^{+0.10}_{-0.15}$, $f_{\rm sed}=1.9^{+1.8}_{-0.8}$. The maximum likelihood sample had $\mathrm{[M/H]}=0.70$, $\mathrm{C/O}=0.55$, and $f_{\rm sed}=1.45$. These parameters are in good agreement with previous work that has found evidence for an effective temperature near $1300\,\mathrm{K}$, a near solar C/O ratio, and tentative evidence of an enriched metallicity \citep[e.g.][]{Zhang2024}.

\par Only the abundance for the enstatite (MgSiO$_3$) cloud converged during the sampling in our atmospheric retrieval; the abundances for the Fe and Na$_2$S clouds reproduced their priors. We find a sedimentation rate $f_{\rm sed}=1.9^{+1.8}_{-0.8}$, an Eddy diffusion rate $\log({K_{\rm zz}/{\rm cm^2s^{-1}}})=8.3\pm1.1$, the width of the log-normal particle size distribution $\sigma_{\rm g}=2.2\pm0.6$, and the abundance of the cloud at the cloud base pressure $X^{c}=(-0.8\pm0.8)\times X_{\rm eq}^c$ (as a fraction of the equilibrium value). We interpret this as strong evidence for a global cloud deck, likely composed of an amorphous silicate, in the planet's photosphere. However, we caution that these wavelengths are not sensitive to the specific spectral features that can distinguish between cloud compositions. Moreover, clouds made of these grains can take on varying size distributions and base pressures, depending on the dominant physical processes (e.g. sedimentation, fragmentation). One should interpret our results only as the unambiguous presence of a cloud that is necessary to reproduce the spectral reddening at short wavelengths.

% \begin{figure}
%     \centering
%     \includegraphics[width=\linewidth]{29Cygb_prt_best.pdf}
%     \caption{Atmospheric modeling of 29~Cyg~b with \texttt{petitRADTRANS}. The best fit model spectrum from the flexible atmospheric retrieval is shown as a blue solid curve. Observations are plotted as in Figure \ref{fig:diamondback}.} 
%     \label{fig:prt}
% \end{figure}

\subsection{Astrometry and orbit}

\par We fit the orbit for 29~Cyg~b using the \texttt{octofitter} package \citep{Thompson2023_octo}. 
The dataset included the relative astrometry of the companion observed with SCExAO \citep{Currie2023_sci, Bovie2025}, GRAVITY \citep{Winterhalder2025}, \added{and the new NIRCam measurement from this work (listed in Table \ref{tab:bka})}, the relative radial velocity of the companion observed with KPIC \citep{Zhang2024}, and the proper motion anomaly of the host star from the Hipparcos Gaia Catalog of Accelerations eDR3 edition \citep{Brandt2021}.
We used \texttt{octofitter} v7 and the Pigeons.jl sampler \citep{Surjanovic2023} for 15 rounds with 24 chains to estimate the posterior on the six Campbell orbital elements, the system parallax (where we again adopted the Gaia DR3 measurement as a prior), and the masses of the two components. We adopted a Gaussian prior on the host star mass $M_\star=1.8\pm0.2\,M_\odot$ \citep{Currie2023_sci}, and a uniform prior on the companion mass ranging between 0.1 and $100\,M_{\rm Jup}$. The orbit solution gives $a_b=14.7\pm0.4\,\mathrm{au}$, $e_b=0.37\pm0.03$, $i_b=160\pm5{^\circ}$, and $M_b=15\pm5\,M_{\rm J}$. The additional coverage along the companion's orbital arc provided by the JWST astrometry improves the precision of the companion's semi-major axis and inclination angle, but otherwise our orbit fit validates previous work \citep{Winterhalder2025, Bovie2025}. %\citet{Bovie2025} notes that adopting the uniform prior on the secondary's mass over the log-normal prior typically adopted by, e.g., \texttt{orvara} \citep{Brandt2021_orv} results in a slightly higher mean value for the distribution on secondary mass ($13$ vs $15\,M_{\rm J}$), with the same uncertainty ($\pm5\,M_{\rm J}$). 

\subsection{Stellar interferometry and apparent spin-orbit alignment}

\added{We investigated the spin-orbit alignment of the system (the angle between the stellar spin axis and the companion orbital plane) by measuring the stellar inclination angle from interferometric observations resolving the oblate host star.}

\added{Observations of 29~Cyg~A were taken with the PAVO instrument at the CHARA Array interferometer \citep{pavo,chara2} on the nights of 2014 April 14, September 15, 18, and 2015 September 6. %\todo{and 2021 data}
We observed 29~Cyg~A with multiple calibrator stars, obtaining 19 calibrated observations. Each calibrated observation yields 24 spectrally dispersed squared visibility measurements. The PAVO instrument uses only two telescopes for each observation. Table \ref{tab:charalog} lists which baseline pair was used for each observing night, as well as the calibrators used, their estimated diameters, and the number of observations made.}

\added{We used the oblate star model (OSM) described in \citet{jones_2015, jones_2016} to analyze the interferometric observations of 29 Cyg A. The OSM calculates the oblate shape and gravity darkening \citep[we assumed the gravity darkening law from][]{ELR_2011} of a potentially rapidly rotating star assuming solid body rotation, and models the interferometric squared visibility at the observed spatial frequencies as well as the star's spectral energy distribution. We describe this model further in Appendix \ref{app:chara}.}

\added{The OSM modeling shows that the rotational inclination of 29 Cyg A is low, with a 95\% confidence interval of 7.8 - 12.6$^\circ$. Given a $v \sin{i}$ of 65 $\pm$ 3 km/s \citep{Royer2007}, this corresponds to 29 Cyg A rotating near or at the critical rate. The full results from the OSM are presented in Table \ref{tab:29CygA_OSM} and discussed in Appendix \ref{app:chara}.}

\par
\begin{figure*}
    \centering
    \gridline{
    \centering
    \fig{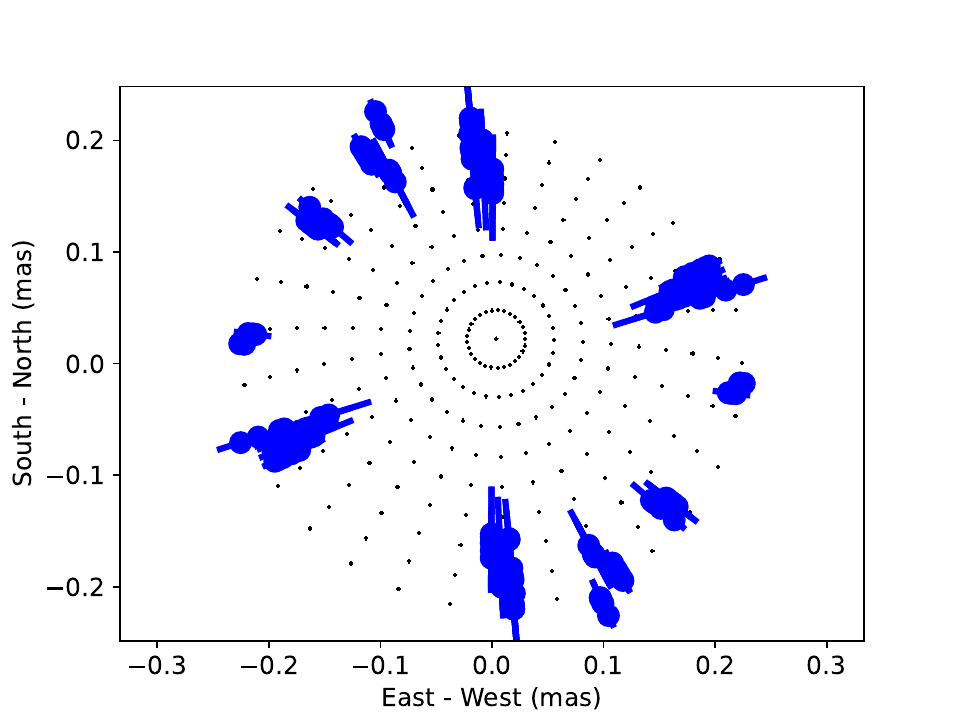}{0.535\textwidth
    }{(a)}
    \fig{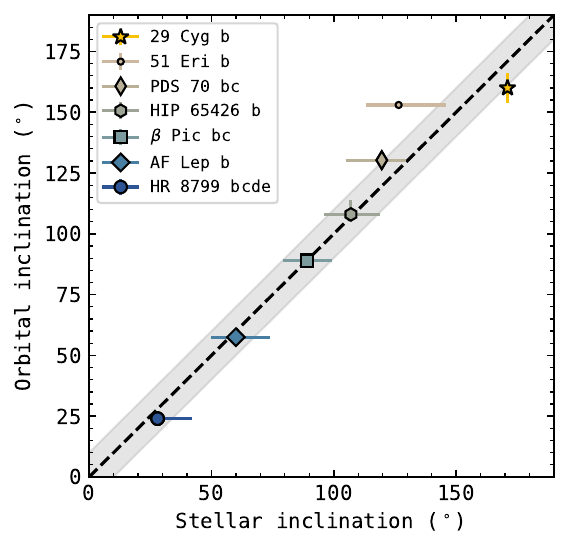}{0.375\textwidth}{(b)}
    }
    \caption{
    % \textbf{(a)} The visual orbit of 29~Cyg~b, including the new relative astrometry from JWST/NIRCam (the top-left-most data point), fit with \texttt{octofitter}. 100 random orbits drawn from the posterior distribution are shown, colored along their path by the mean anomaly of the orbit (the fraction of the orbit completed since periastron passage). 
    % \textbf{(b)}
    \textbf{(a)} \added{The surface of the best model fit to the CHARA/PAVO observations of 29~Cygni~A. Dots represent the grid of co-latitudes and longitudes facing the observer. Blue circles and errorbars represent radii fitted to each visibility measurement at a given baseline orientation. The observations are duplicated at 180$^\circ$.}
    % \added{Apparent rotation period for 29~Cyg~A from TESS photometry. A Lomb-Scargle periodogram of TESS Sectors 14, 15, 41, 54, 55, 75, and 82 is plotted in blue, and the 1\% False Alarm Probability level for this periodogram as a dashed black line. High frequency pulsations are apparent. An inset panel shows the low frequencies between 1 and 10 cycles/day, where a 0.72 d period lies above the False Alarm threshold. Another inset panel illustrates a Gaussian fit to this period, which we tentatively adopt as the stellar rotation period.}
    \textbf{(b)} Spin-orbit alignment trend for planetary mass, directly imaged companions with orbits $a<100\,\mathrm{au}$. This figure follows from \citet[][their Figure 7]{Sepulveda2024}, but with orbits updated from the literature and including our new measurement of 29~Cyg~b (yellow star). References for stellar and planetary inclination angles are, respectively: HR~8799~bcde \citep{Wang2018, Sepulveda2023}, PDS~70~bc \citep{Trevascus2025, Bowler2023}, $\beta$~Pic~bc \citep{Lacour2021, Zwintz2019}, HIP~65426~b \citep{Blunt2023, Sepulveda2024}, AF~Lep~b \citep{Balmer2025a, Zhang2023}, 51~Eri~b \citep[Balmer et al. in prep.;][]{Bowler2023} and 29~Cyg~b (this work). 
    }
    \label{fig:incs}
\end{figure*}

 Assuming the pro-grade configuration between the stellar rotation axis and the orbit (that is, selecting for the $171.0_{-1.3}^{+0.7}\,{^\circ}$ peak), the stellar spin and planetary orbit are aligned to within $\Delta i=12\pm6^\circ$. At the very least, it can be stated that there is no strong evidence for spin-orbit misalignment in this system. This result follows a growing trend for planetary mass directly imaged companions, which is shown in Figure \ref{fig:incs}. This result contrasts with the distribution of angles for massive and widely separated brown dwarf companions to cool stars in \citet{Bowler2023}, who show that there is no evidence for spin-orbit alignment between higher mass ratio brown dwarf companions and their hosts.

\section{Discussion and Conclusions}

Since the discovery of the first directly imaged companions orbiting on planetary scales around early type stars---like HR~8799~bcde \citep{Marois2008} and $\beta$~Pic~b \citep{Lagrange2009}--- planet formation models have been constructed to explain the formation of super-Jupiters in protoplanetary disks. It has become apparent that global core accretion models in massive disks around early type stars can produce super-Jupiter planets matching observations up to and exceeding the deuterium burning limit \citep[e.g.][figure 16 for $\beta$~Pic~b]{Mordasini2015}. Disk fragmentation models can create companions with low mass ratios, but are far more efficient at creating stellar and brown dwarf mass companions than planetary mass companions \citep{Kratter2010, Kratter2016, Forgan2018}. This pathway can in principle produce giant planets when disk parameters are well tuned, but these appear to be quickly ejected \citep{Boss2023} or tidally disrupted \citep{Zhu2012}. There is also growing evidence that low mass ratio ($\lesssim0.01$) companions, orbiting their stars  within the typical outer scale of a protoplanetary disk ($\sim 100\,\mathrm{au}$) exhibit enhanced metallicities when compared to widely separated, high mass ratio planetary mass companions \citep{Xuan2024}. This is consistent with significant (and therefore rapid) solid accretion \citep{Wang2025}, and seems to indicate that the former are the outcome of a bottom-up process within the disk and the latter stem from top-down fragmentation of gas. 

\par Dynamical evidence for a split between orbital eccentricity distributions \citep{Bowler2020, Nagpal2023, DoO2023} and spin-orbit alignment \citep{Bowler2023, Biddle2025} between these two populations supports this interpretation. Recent results from the California Legacy Survey suggest that this differentiation from planets to brown dwarfs is muddied. There doesn't appear to be a distinct split in the distribution of semi-major axes between the two populations \citep{VanZandt2025}, and the change from predominately low eccentricities to isotopically distributed eccentricities is gradual \citep{Gilbert2025}. The change point in the distributions of stellar metallicities between the two populations in this updated sample occurs near $\sim30\,M_{\rm J}$ \citep{Giacalone2025}. This motivates using atmospheric and/or bulk abundances as more reliable proxies for formation, for individual investigations.

So, it seems clear that core/pebble accretion can create massive super-Jovian planets between $2-10\,M_{\rm J}$ \citep{Burn2024}, and that these planets can be distinguished (albeit with great observational effort) from brown dwarf companions. Observational evidence for planets with masses exceeding the deuterium burning limit has been less apparent. \citet{Currie2023_sci} argued that the mass ratio and orbital separation for 29~Cyg~b are indicative of a similarity with lower mass planets. The recent discovery of a second companion in the HD~206893 system also provides a unique insight into this question \citep{Hinkley2023}. In this system, orbiting an early type star, are two directly imaged companions that are relatively co-planar with a debris disk, one at the deuterium burning limit $M_c\sim12\,M_{\rm J}$ and another exceeding it $M_b\sim20\,M_{\rm J}$. HD~206893~c appears to exhibit enhanced luminosity due to deuterium burning, given the system's age \citep[][their figure 5]{Hinkley2023}. The metallicities of these two companions have been challenging to measure due to the strong extinction from clouds in their atmospheres that have reddened their near infrared spectrophotometry significantly, and confounded atmospheric modeling \citep{Ward-Duong2021, Kammerer2021, Sappey2025}.

In this context, the apparent atmospheric enrichment and spin-orbit alignment of 29~Cyg~b are additional clues that, around early type stars, planets can form and accrete significant fractions of metals while growing to masses at or exceeding the deuterium burning limit. Assuming effectively solar abundances for the host \citep[the LTE and non-LTE abundance of C, N, O, S, Ca for 29~Cyg from ][range from -0.5 to -0.1 dex]{Paunzen1999b, Kamp2001} and the atmospheric metallicity from our \texttt{petitRADTRANS} retrieval ($\mathrm{[M/H]}=0.42\pm0.25$) with the conversion in \citet{Thorngren2019}, we find $Z_{\rm pl}/Z_\star=3\pm2$. This enrichment agrees within uncertainties with the trend established by the densities of warm transiting planets and fit in \citet{Thorngren2016} and revised by \citet{Chachan2025}. Their trend predicts $Z_{\rm pl}/Z_\star=3.3\pm0.5$.
To estimate the order of magnitude metal mass fraction of 29~Cyg~b, we follow \citet{Wang2025}, taking their equation 4 \citep[that is effectively equation 2 in][]{Oberg2011}. We assume, following the discussion in \S4.1 of \citet{Wang2025}, that $f_{sg}=0.01$, $M_g=15\,M_{\rm J}$, $\log\alpha_M=0.5-0.75$, $\eta=0.75$, as well as formation outside the CO ice line so that $f_{C,s}$ and $f_{O,s}$ are unity. We find an \added{accreted solid mass of $M'_s=170^{+130}_{-95}\,M_{\oplus}$}, similar to the estimate in \citet{Wang2025} for the planets AF~Lep~b and $\beta$~Pic~b. 

\added{How can such a large planet accrete this much metal-rich material? Major-mergers have been suggested as a way for super-Jupiters to achieve large metal enrichments, as concurrent gas accretion and mergers at $\sim10\,\mathrm{au}$ are able to reproduce both the behavior and scatter in the observed mass-metallicity trend for transiting planets, which exhibits a number of metal rich super-Jupiters akin to directly imaged planets \citep{Ginzburg2020}. It is also possible that such large enrichment can be generated by the formation of a vortex within the protoplanetary disk \citep[e.g.][]{Klahr2006, Bae2015, Raettig2021} that funnels solid material into the circumplanetary environment, enhancing the \added{dust-to-gas ratio} locally. Formation interior to an ice-line and subsequent pebble drift can also enrich the gas phase of an embedded planet \citep{Schneider2021a, Schneider2021b}. The recently proposed Dust Recycling and Icy Volatile Enhancement (DRIVE) mechanism \citep{VanClepper2025}, where the stirring of small grains trapped in meridional flows around a forming planet can repeatedly sublimate CO ice into the circumplanetary environment, could generate enhanced planetary metallicities at wide separations ($>10\,\mathrm{au}$), where previously proposed mechanisms were inefficient.} For the analogous HR~8799 system (super-Jovian planets orbiting around a $\lambda$~Boo A-type star with a far-IR excess), \citet{Helled2010} find that formation by disk-instability leaves the planets at wide separations unable to efficiently capture enough solids to appreciably raise their metallicity above the stellar value. This is because there is a limited timescale before the gravitationally unstable core reaches the temperatures $\sim2000\,\mathrm{K}$ where H$_2$ disassociates and efficient planetesimal accretion into the core ends. The large atmospheric enrichment our observations indicate could therefore be evidence that 29~Cyg~b formed from the bottom-up, by the coagulation of solids, and/or that an (unknown) process fed metal-rich material onto the planet over the course of its gas accretion phase. 

\par In the near future, measurements of the CO$_2$ abundance of other directly imaged planets with JWST (from both our team's ongoing GO Program 6905, and other direct imaging/spectroscopy efforts) should provide updated constraints on their bulk metallicities and enable the measurement of trends across mass, orbital parameters, and atmospheric properties. These can be used to revise our understanding of the formation mechanisms and timescales of giant planets. Higher resolution measurements of 29~Cyg~b, with, for example JWST/NIRSpec forward modeling \citep{Ruffio2024}, can leverage our precise, space based photometry to interpret the molecular, even isotopic signatures in the planet's atmosphere and investigate the specific formation location of this massive planet in detail. Of particular interest will be the refractory to volatile abundance ratios \citep[e.g.][]{Lothringer2021, Chachan2023} that encode information about the type of material (pebbles, planetesimals, metal-rich gas, etc) accreted by the companion during its formation. NIRSpec should be able to better constrain the presence (or absence) of the 3.3\um absorption feature due to methane, and therefore the \added{strength} of the vertical transport in the atmosphere. This will help better refine the bulk abundance measurement, and in turn, the inference on total metal mass in the near future.

\begin{acknowledgments}
We thank the anonymous reviewer whose constructive report improved this letter.
We thank David Golimowski and Tony Roman, the instrument scientist and primary support contact at STScI for JWST GO 6905.
W.O.B. thanks Dan Huber for the fruitful discussion regarding the TESS lightcurve of 29~Cyg~A, John Monnier and Matthew De Furio for their help concerning CHARA interferometric observations of 29~Cyg~A, and Paul Molli\`ere for the discussion of CO$_2$ quenching.
This work benefited from the 2025 Exoplanet Summer Program in the Other Worlds Laboratory (OWL) at the University of California, Santa Cruz, a program funded by the Heising-Simons Foundation and NASA. Atmospheric modeling was carried out by W.O.B. at the Advanced Research Computing at Hopkins (ARCH) core facility (rockfish.jhu.edu), which is supported by the National Science Foundation (NSF) grant number OAC1920103, as well as at the STScI Science Cluster.
This work makes use of observations obtained with the Georgia State University Center for High Angular Resolution Astronomy Array at Mount Wilson Observatory. The CHARA Array is supported by the National Science Foundation under Grant No. AST-2034336 and AST-2407956. Institutional support has been provided from the GSU College of Arts and Sciences and the GSU Office of the Provost and Office of the Vice President for Research and Economic Development.

\end{acknowledgments}

\begin{contribution}

W.O.B. and L.A.P. devised and proposed the observations, and led the collaboration. W.O.B. wrote the manuscript. W.O.B., A.M., and E.B. completed the data reduction, starlight subtraction, and source forward modeling. J.J. observed and modeled the stellar oblateness. K.L. conducted the position dependent deconvolution of the NIRCam images. A.M. and W.O.B. completed the orbit fitting. W.O.B. and K.M. completed the atmospheric model comparison. W.O.B. compiled the spin-orbit inclination angles from the literature. All other authors contributed feedback to the initial proposal, observation planning, and manuscript writing.

\end{contribution}

\facilities{JWST(NIRCam), TESS}

\software{\texttt{NumPy} \citep{numpy},
\texttt{SciPy} \citep{scipy},
\texttt{Matplotlib} \citep{matplotlib},
\texttt{Astropy} \citep{astropy2013, astropy2018, astropy2022},
\texttt{Astroquery} \citep{astroquery},
\texttt{pyKLIP} \citep{Wang2015},
\texttt{spaceKLIP} \citep{Kammerer2022},
\texttt{winnie} \citep{Lawson2023},
\texttt{species} \citep{Stolker2020},
\texttt{petitRADTRANS} \citep{Molliere2019},
\texttt{easyCHEM} \citep{Lei2024},
\texttt{octofitter} \citep{Thompson2023_octo},
}

\newpage
\appendix
\onecolumngrid
\section{Data reduction} \label{app:data}

\par We observed the target star at two telescope roll angle positions, separated by 7$^\circ$, and the dedicated reference star 36~Cyg using the \texttt{5-POINT-BAR} small grid dither pattern, to facilitate Angular and Reference Differential Imaging \citep[ADI+RDI; for a review, see][]{Follette2023}. Target acquisition on the coronagraphic neutral density squares used the F335M filter, the \texttt{SHALLOW4} readout pattern with 17 groups per integration (33 groups per integration for the reference star), and astrometric confirmation images on the target used the same filter in the \texttt{RAPID} readout pattern with 4 groups per integration. Observations at each pointing were taken through the F410M/F430M/F460M filters in the Long Wavelength (LW) detector channel, and through the F210M filter in the Short Wavelength (SW) detector channel. On the first roll, the detector was read out in the \texttt{SHALLOW4} pattern with 10 groups per integration, for 4/8/8 integrations per exposure, on the second roll with 8/16/12 integrations per exposure, and on the reference star, 4/6/5 integrations per exposure across the five dither positions (for a total of 20/30/25 integrations). Slightly more exposure time was dedicated to the F430M filter than the others, prioritizing the measurement of the CO$_2$ feature, because the F410M filter has a wider bandpass than F430M or F460M, and because the F460M filter has a ground based equivalent.

\begin{figure}
    \centering
    \includegraphics[width=\linewidth]{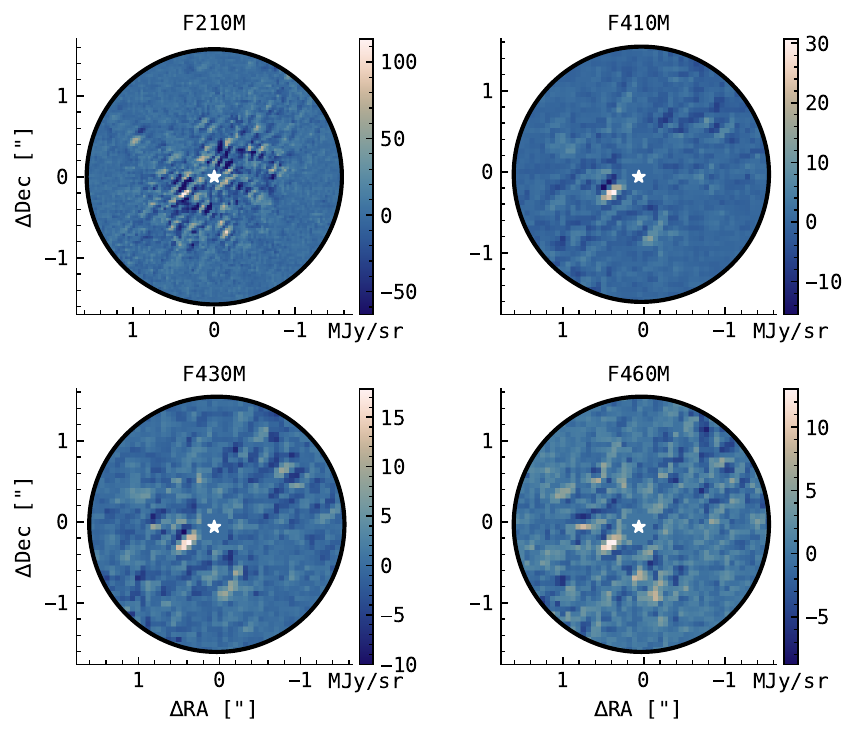}
    \caption{Coronagraphic images of 29~Cyg~b with the JWST/NIRCam \texttt{LWBAR} coronagraph at the narrow offset position in the F210M, F410M, F430M, and F460M filters, respectively.}
    \label{fig:detections}
\end{figure}

\par The images were acquired from the Barbara A. Mikulski Archive for Space Telescopes (MAST)\footnote{Dataset at\dataset[doi:10.17909/n88j-z828]{https://dx.doi.org/10.17909/n88j-z828}}. We reduced the data with the \texttt{jwst} pipeline \citep{Bushouse2023} wrapped by the \texttt{spaceKLIP} package \citep{Kammerer2022, Carter2023}, following the steps and modifications outlined in \citet{Balmer2025b} that are appropriate for \texttt{MASKLWB}/narrow data. The data reduction used \texttt{jwst} version \added{1.20.2}, CRDS version 12.1.5 and the \texttt{jwst\_1464.pmap} CRDS context\footnote{\url{https://jwst-crds.stsci.edu/display\_context\_history/}}. Starlight subtraction, contrast estimation, and source forward modeling was done using \texttt{pyKLIP} \citep{Wang2015} via \texttt{spaceKLIP}, again following previous work \citep{Franson2024, Balmer2025b, BardalezGagliuffi2025}. The companion was detected in all filters. During image alignment, it became apparent that stochastic target acquisition uncertainties resulted in the 36~Cyg reference observations landing behind the opposite edge of the coronagraph compared to the 29~Cyg target observations, resulting in significant correlated residual noise after starlight subtraction\footnote{These target acquisition uncertainties were nevertheless well within the expected performance for NIRCam coronagraphy, less than 15 milliarcseconds in magnitude.}. We therefore included previous PSF reference observations of HD~30562 (originally used as reference for 51~Eri) and HD~220657 (originally used as reference for HR~8799) that were taken in the same filters and coronagraph-offset mode. The inclusion of these reference observations improved the signal to noise of the detections, because they provided a denser sampling of the position dependent PSF near the location of the target star behind the coronagraph \citep{Soummer2014}. Figure \ref{fig:detections} shows the starlight subtracted images of 29~Cyg~b, using KLIP with a single annulus with a radius of 1\farcs0 centered on the star and 200 principle components (we proceed with this starlight subtraction in all subsequent analyses). 

We used \texttt{STPSF} \citep{Perrin2025} version 2.1\footnote{In particular, we used the development version of \texttt{STPSF} git commit \href{https://github.com/spacetelescope/stpsf/commit/2645d4c97d960c63b7dd6315446a514559f96a40}{2645d4c}, which incorporates a change in the orientation of the applied optical path difference map (measured from wavefront sensing observations) used during the generation of NIRCam coronagraphic off-axis PSFs, due to an apparent coordinate system inconsistency between benchmark Zemax models and earlier versions of \texttt{STPSF}. This update appears to resolve a long-standing residual pattern observed in companion forward modeling with \texttt{STPSF}/\texttt{spaceKLIP}, visible in, e.g. \citep{Carter2023} Appendix C. This residual and the associated inaccuracy in derived photometry predominately affects high SNR PSFs.} to construct off-axis PSF models based on near-in-time optical path difference measurements \citep{Perrin2012, Perrin2014}. We used these PSF models to measure the planet astrometry and photometry through the KLIP-FM algorithm \citep{Pueyo2016}, accounting for coronagraphic and algorithmic throughput following previous work with spaceKLIP \citep{Carter2023}. The astrometry for the 2025-09-01 observation was taken to be the weighted sum of the astrometry measured for each detection, with uncertainty on the weighted average propagated with a systematic uncertainty on the center position of the host star of 6.3 mas following \citet{Carter2023}. Table \ref{tab:bka} records the photometry (and associated signal-to-noise ratio) of 29~Cyg~b in each filter, as well as the combined astrometry for the observing sequence. 

We created a deconvolved false-color image of 29~Cyg~b. The deconvolution followed \citet{Balmer2025b}, Appendix B; we used the \texttt{winnie}\footnote{\url{https://github.com/kdlawson/Winnie}} package \citep{Lawson2023, Lawson2024} to process the RDI subtracted images in each LW filter with 30 iterations of a spatially-dependent, coronagraphic Richardson-Lucy deconvolution algorithm leveraging the \texttt{STPSF} forward models from the photometric forward modeling step. These were then re-convolved with a Gaussian with a full-width half-max of 0.75 pixels, and normalized to the forward model photometry in each filter. 

\begin{deluxetable*}{cCCCCC}
\tablewidth{\textwidth}
\tablecaption{Relative astrometry and photometry for 29~Cyg~b. \label{tab:bka}}
\tablehead{
\colhead{Filter} & \colhead{$\lambda_{\rm cen}$} & \colhead{$\mathrm{SNR}_{\rm bf, \mathcal{F}}$} & \colhead{$\Delta\mathrm{RA}$} & \colhead{$\Delta\mathrm{Dec}$} & \colhead{Flux density} \\
\colhead{} & \colhead{[\um]} & \colhead{} & \colhead{[mas]} & \colhead{[mas]} & \colhead{[\textmu Jy]}
}
\startdata
F210M & 2.097 & 5 & \nodata        & \nodata      & 390\pm25      \\ 
F410M & 4.084 & 12 & \cdots & \cdots &  560\pm15 \\ 
F430M & 4.283 & 7 & \cdots & \cdots &  378\pm20 \\ 
F460M & 4.631 & 6 & \cdots & \cdots &  322\pm18 \\
all & \cdots & \cdots & 336.6\pm1.7 & -161.8\pm1.5 & \cdots \\
% \hline 
\enddata
22\tablecomments{$\mathrm{SNR}_{\rm bf, \mathcal{F}}$ as defined in \citet{Golomb2021} is the standard deviation of the nearby pixels compared to the peak flux on the planet; it is not calibrated to account for, e.g. small sample statistics or the algorithmic throughput. $\Delta\mathrm{RA}$, $\Delta\mathrm{Dec}$ are the position measurements relative to the central star, and flux density is the brightness of the source, accounting for algorithmic and coronagraphic throughput, measured as described in \S\ref{sec:obs}.}
\end{deluxetable*}

\onecolumngrid 
\section{Atmospheric modeling} \label{app:atmo}

We used \texttt{petitRADTRANS} v3.2 to model the SED of 29~Cyg~b using a flexible, parameteric atmospheric model. The atmospheric retrieval used a correlated-k treatment for opacities and included line species for H$_2$O \citep{ExoMol_H2O}, CO \citep{rothman_hitemp_2010}, CH$_4$ \citep{ExoMol_CH4}, CO$_2$ \citep{ExoMol_CO2}, NH$_{3}$ \citep{ExoMol_NH3}, HCN \citep{ExoMol_HCN}, H$_{2}$S \citep{ExoMol_H2S}, PH$_{3}$ \citep{exomol_ph3}, FeH \citep{wende_feh_2010}, Na \citep{allard_new_2019}, K \citep{allard_k-h_2016}, SiO \citep{exomol_sio}, TiO \citep{exomol_tio}, and VO \citep{exomol_vo}, Rayleigh scattering opacities for H$_2$ and He \citep{Dalgarno1962, Chan1965}, collision induced absorption (CIA) of H$_2$ and He \citep{Borysow1988, Borysow1989, Borysow2001, Richard2012}, and crystalline (DHS irregular) opacities for iron \citep[Fe,][]{Henning1996}, amorphous (DHS irregular) enstatite \citep[MgSiO$_3$,][]{Scott1996, Jaeger1998}, and crystalline (DHS irregular) sodium sulfide \citep[Na$_2$S,][]{Morley2012} grains that form clouds. These opacities are available from the ExoMol database \citep{Chubb2021}, in the ``petitRADTRANS" format.\footnote{\href{https://www.exomol.com/data/data-types/opacity/}{www.exomol.com}}.

For the P-T structure, we adopted the temperature gradient parameterization introduced by \citet{Zhang2023}, where the $d\ln T/d\ln P$ of the P-T curve at a given pressure is sampled instead of the temperature at a given pressure. This allows the user to set prior constraints on the gradient based on ``self-consistent" radiative convective equilibrium models. We followed the implementation in \citet{Zhang2023} exactly, and sampled at six pressure nodes spaced logarithmically between $1000-10^{-3}\,\mathrm{Bar}$, with the normally distributed priors listed in their Equation 3 \citep[and in Table 6 in][]{Nasedkin2024}, derived from the RCE models from \citet{Phillips2020, Marley2021, Karalidi2021, Mukherjee2022, Lacy2023}. Above $10^{-3}\,\mathrm{Bar}$, the atmosphere is set isothermal to the temperature at $10^{-3}\,\mathrm{Bar}$ (this is acceptable because there is effectively no contribution from these pressures to the retrieval given the resolution and wavelength range of the data).

The retrieval relied on the equilibrium chemistry parameterization available in \texttt{petitRADTRANS} and described in \citet{Molliere2020, Nasedkin2024} for the retrieval module. The equilibrium chemical abundances of parameters are set as a function of pressure, temperature, bulk metallicity ([M/H]), and carbon to oxygen ratio (C/O), and are determined for a random sample of P, T, [M/H], and C/O by interpolating on a table computed with \texttt{easyCHEM} \citep{Lei2024}. For clouds, we used the \texttt{EddySed} parameterization described first in \citet[][]{Ackerman2001}, implemented in \texttt{petitRADTRANS} following \citet{Nasedkin2024}.

\par \added{The impact of vertical mixing induced disequilibrium chemistry was parameterized by retrieving a ``quench" pressure, a pressure beyond which the abundances of H$_2$O, CO, and CH$_4$ are set constant (in a self consistent model, this would be the point where the mixing timescale equals the chemical timescale, here $P_{\rm quench}$ is a free parameter ranging between $10^3$ and $10^{-6}$ Bar. Importantly, in this retrieval set-up, the quench approximation is not applied to any trace species (such as CO$_2$), so these species follow their chemical equilibrium values for a given P, T, [M/H], and C/O. CO$_2$ does quench off the quenched CO abundance \citep{zahnle2014, Wogan2025}, which in colder atmospheres of Y-type brown dwarfs impacts the observed spectrum significantly \citep{Beiler2024}, but the quenching of CO$_2$ occurs above the pressures and temperatures probed by our observations for 29~Cyg~b.}

\begin{table}[t!]
\centering
{ \footnotesize
\caption{Priors and posterior summary statistics for the 29~Cyg~b \texttt{petitRADTRANS} atmospheric retrieval. $\mathcal{U}$ stands for a uniform distribution, with the two parameters being the range boundaries. $\mathcal{N}$ stands for a Gaussian distribution, with the two parameters being the mean and standard deviation.}
\begin{tabular}{llLL}
\hline
Parameter & Prior & \rm Median~ (\pm 1\,\sigma) & \rm Max.~Likelihood \\ \hline \hline

\textit{Bulk properties} & & & \\

Parallax [mas] & $\mathcal{N}(24.55,0.09)$ & 24.55\pm0.09 & 24.55 \\
Radius [$R_{\rm J}$] & $\mathcal{N}(1.3,0.1)$ & 1.17\pm0.05 & 1.18 \\
Mass [$M_{\rm J}$] & $\mathcal{N}(15,5)$ & 14.3\pm2 & 9.8 \\

\hline

\textit{P-T profile} & & & \\

$T_{\rm P=10^3\,bar} [\rm K]$ & $\mathcal{U}(2000,20000)$ & 8690^{+1240}_{-1080} & 7860 \\
$(d\ln{}T/d\ln{}P)_1$ & $\mathcal{N}(0.25, 0.025)$ & 0.25\pm0.02 & 0.20 \\
$(d\ln{}T/d\ln{}P)_2$ & $\mathcal{N}(0.25, 0.045)$ & 0.25\pm0.03 & 0.21 \\
$(d\ln{}T/d\ln{}P)_3$ & $\mathcal{N}(0.26, 0.05)$ & 0.26\pm0.03 & 0.28 \\
$(d\ln{}T/d\ln{}P)_4$ & $\mathcal{N}(0.2,0.05)$ & 0.13\pm0.03 & 0.13 \\
$(d\ln{}T/d\ln{}P)_5$ & $\mathcal{N}(0.12 , 0.045)$ & 0.09\pm0.03 & 0.7 \\
$(d\ln{}T/d\ln{}P)_6$ & $\mathcal{N}(0.07, 0.07)$ & 0.07\pm0.06 & 0.0 \\
\hline

\textit{Chemistry} & & & \\
$\rm C/O$ & $\mathcal{U}(0.1,1.0)$ & 0.45^{+0.09}_{-0.16} & 0.55 \\
$\rm [Fe/H]$ & $\mathcal{U}(-0.5,2.0)$ & 0.42^{+0.26}_{-0.34} & 0.69 \\
${\rm log}(P_{\rm quench}/{\rm bar})$ & $\mathcal{U}(-6,3)$ & -3.2\pm1.7 & -4.4 \\ 
\hline

\textit{Clouds} & & & \\
$\log({K_{\rm zz}/{\rm cm^2s^{-1}}})$ & $\mathcal{U}(5,13)$ & 10.1^{+1.2}_{-1.5} & 7.33 \\
$\sigma_{g}$ & $\mathcal{U}(1.05,3.0)$ & 2.2\pm0.6 & 2.3 \\
% $f_{\rm sed,\,MgSiO_3 (am)}$ & $\mathcal{U}(0.1,10.0)$ & 1.1^{1.0}_{-0.5} & \\
% $f_{\rm sed,\,Mg_2SiO_4}$ & $\mathcal{U}(0.1,10.0)$ & 5.7^{+2.7}_{-3.0} & \\
% $f_{\rm sed,\,Fe}$ & $\mathcal{U}(0.1,10.0)$ & 5.5\pm2.9 & \\
$f_{\rm sed}$ & $\mathcal{U}(0.1,10.0)$ & 1.9^{+1.8}_{-0.8} & 1.46 \\
$\log(X^{\rm Na_2S}_0/X^{\rm Na_2S}_{\rm eq})$ & $\mathcal{U}(-3.5,1.0)$ & -2.0^{+1.7}_{-1.8} & -1.9 \\
$\log(X^{\rm MgSiO_3}_0/X^{\rm MgSiO_3}_{\rm eq})$ & $\mathcal{U}(-3.5,1.0)$ & -0.75\pm0.85 & -2.0 \\
$\log(X^{\rm Fe}_0/X^{\rm Fe}_{\rm eq})$ & $\mathcal{U}(-3.5,1.0)$ & -3.2\pm1.2 & -4.5 \\
\hline

\end{tabular}
}
\label{tab:prt_prior}
\end{table}

The parameters and their adopted priors are listed in Table \ref{tab:prt_prior}. The sampling was performed using the \texttt{Multinest} nested sampling algorithm \citep{Feroz2008, Feroz2009, Feroz2019}, with the \texttt{pyMultinest} package \citep{Buchner2014}. The sampler was run with 960 live points on 64 cores until Multinest determined the sampler had reached convergence at a global log evidence of $\ln{Z}=9350\pm2$ and an acceptance rate of 0.015, after 1,100,000 total likelihood evaluations.

\section{Interferometric modeling of 29~Cyg~A} \label{app:chara}

Table \ref{tab:charalog} records the observing log for the CHARA/PAVO observations of 29~Cyg~A considered in this work.

\begin{table*}[h]
	\begin{center}
	\caption{Log of CHARA/PAVO Observations for 29 Cyg A. \label{tab:charalog}}
	\begin{tabular}{ccccc}
	\hline
	Cal HD & Cal Diameter (mas) & Baseline & \# Observations & Date \\
	\hline \hline
	188892 & 0.138 $\pm$ 0.014 & S2-W2 & 3 & 2014 Apr 14 \\
	188892 & 0.138 $\pm$ 0.014 & W2-E2 & 7 & 2014 Sep 15 \\
    193369 & 0.249 $\pm$ 0.025 & W2-E2 & 3 & 2014 Sep 15 \\
    197392 & 0.205 $\pm$ 0.021 & S2-W1 & 5 & 2014 Sep 18 \\
    194789 & 0.111 $\pm$ 0.011 & E2-W1 & 1 & 2015 Sep 6 \\
	\hline
	\end{tabular}
	\end{center}
\end{table*}

The OSM determines the oblate shape and gravity darkening of a potentially rapidly rotating star assuming solid body rotation, and models the interferometric squared visibility at the observed spatial frequencies as well as the star's spectral energy distribution. The these modeled values are compared to the observed squared visibilities, as well as photometric fluxes taken from the literature\footnote{Flux measurements were taken from \cite{mermilliod_1991} - UBV, \cite{hauck_1997} - uvby, \cite{ducati_2002} - JHK, \cite{thompson_1978} - F2740, F2365, F1965, F1565}. There are two significant differences from the version of the model presented in that work. Firstly, the model presented here used a Markov Chain Monte Carlo algorithm to determine the posterior probability distribution function (PDF) of the modeled parameters, using the $\chi^2$ of the model as its likelihood, instead of using a $\chi^2$ minimization routine. Secondly, the $v \sin{i}$ was treated as a gaussian prior in the MCMC fit, using the value from \citet{Royer2007}.

Briefly, the OSM has five free parameters and three fixed parameters. The free parameters are equatorial radius ($R_\mathrm{e}$); equatorial rotational velocity ($V_\mathrm{e}$); inclination of the polar axis relative to our line of sight ($i$); polar temperature ($T_\mathrm{p}$); and position angle ($\psi$). Both $i$ and $\psi$ have a 180$^\circ$ ambiguity. The fixed parameters are mass ($M_*$); the gravity darkening coefficient ($\beta$); and stellar parallax ($\pi_\mathrm{plx}$). The value for $\pi_\mathrm{plx}$ is that measured by \emph{Hipparcos} \citep{vanleeuwen_2007}. For $\beta$, we use the gravity darkening law of \cite{ELR_2011}. The $M_*$ that we adopt is a result of fitting the initial model parameters to evolution models. The OSM is not very sensitive to errors in the mass used \citep{jones_2015}. In order to determine that the Markov chains have converged and are sampling the region of parameter space that corresponds with the maximum of the joint PDF, three chains with different initial parameters were ran until 10,000 iterations were accepted by the sampling algorithm after the minimum value of the $\chi^2$ for that chain was determined. Thus, the final PDFs are made up of 30,000 total iterations.

The OSM modeling shows that the rotational inclination of 29 Cyg A is low, with a 95\% confidence interval of 7.8 - 12.6$^\circ$. 
Given $v \sin{i}=65\pm3$ km/s \citep{royer_2007}, this corresponds to a rotation speed at or near the critical rate. The full results from the OSM are presented in Table \ref{tab:29CygA_OSM}. With a 95 \% confidence interval of 0.35 - 0.39 mas, the angular diameter of 29 Cyg A is too small for the interferometric observations to constrain all five of the OSM's free parameters by themselves. At this size, the visibilities are not sensitive to limb- or gravity darkening, but are sensitive to the angular diameter and oblateness. Together with the parallax, the interferometry is sufficient to constrain $R_\mathrm{e}$ and demonstrate that the apparent shape is too symmetric to constrain $\psi$. The photometric fluxes provide important constraints on the temperature profile, and together with the squared visibilities, they constrain $V_\mathrm{e}$ and $i$.

\begin{table*}
\begin{center}
	\caption{Properties of 29 Cyg A using the Oblate Star Model. \label{tab:29CygA_OSM}}
	\begin{tabular}{ccccc}
		\hline \hline
        \multicolumn{5}{c}{Modeled Parameters} \\
		\tableline
        & Median & 68\% & 95\% & Minimum $\chi^2$ \\
		\tableline
		Equatorial Radius, $R_\mathrm{e}$ (R$_{\sun}$) & 2.32 & 2.09 - 2.72 & 2.03 - 3.20 & 2.24 \\
		Equatorial Velocity, $V_\mathrm{e}$ ($\mathrm{km~s^{-1}}$) & 414 & 362 - 436 & 302 - 456 & 410 \\
		Stellar Inclination, $i$ ($^\circ$) & 9.1 & 8.4 - 10.5 & 7.8 - 12.6 & 8.2 \\
		Polar Temperature, $T_\mathrm{p}$ (K) & 8800 & 8754 - 8838 & 8697 - 8875 & 8801 \\
		Polar Position Angle, $\psi$ ($^{\circ}$) & $-$4 & $-$53 - $+$47 & $-$83 - $+$83 & $-$10 \\
		\tableline
        \multicolumn{5}{c}{Derived Parameters} \\
		\tableline
		Gravity Darkening, $\beta$ & 0.16 & 0.14 - 0.18 & 0.13 - 0.19 & 0.14 \\
		Angular Rotation Rate, $\omega$ & 0.99 & $>$ 0.96 & $>$ 0.92 & 0.99 \\
		Polar Radius, $R_\mathrm{p}$ (R$_{\sun}$) & 1.50 & 1.45 - 1.57 & 1.42 - 1.65 & 1.45 \\
		Average Radius, $R_\mathrm{avg}$ (R$_{\sun}$) & 1.70 & 1.66 - 1.76 & 1.62 - 1.81 & 1.67 \\
		Average Angular Diameter, $\theta_\mathrm{avg}$ (mas) & 0.37 & 0.36 - 0.38 & 0.35 - 0.39 & 0.36 \\
		Equatorial Temperature, $T_\mathrm{e}$ (K) & 6210 & 5735 - 6599 & 5127 - 6965 & 5634 \\
		Average Temperature, $T_\mathrm{avg}$ (K) & 8062 & 8028 - 8098 & 7992 - 8138 & 8056 \\
		Polar Surface Gravity, $\log(g_\mathrm{p})$ (cgs) & 4.34 & 4.30 - 4.37 & 4.26 - 4.39 & 4.37 \\
		Average Surface Gravity, $\log(g_\mathrm{avg})$ (cgs) & 4.15 & 4.12 - 4.17 & 4.11 - 4.19 & 4.17 \\
		Equatorial Surface Gravity, $\log(g_\mathrm{e})$ (cgs) & 3.37 & 3.04 - 3.60 & 2.52 - 3.73 & 2.97 \\
		$v\sin i$ ($\mathrm{km~s^{-1}}$) & 65 & 62 - 68 & 58 - 71 & 58.5 \\
		Total Luminosity, $L_\mathrm{tot}$ (L$_{\sun}$) & 10.0 & 9.7 - 10.9 & 9.4 - 11.7 & 9.6 \\
		Apparent Luminosity, $L_\mathrm{app}$ (L$_{\sun}$) & 15.9 & 15.7 - 16.1 & 15.4 - 16.4 & 15.9 \\
		\hline
	\end{tabular}
\end{center}
\end{table*}

% \newpage
\bibliography{minibar}{}
\bibliographystyle{aasjournalv7}

\end{document}